\documentclass[prx,twocolumn,amsmath,amssymb]{revtex4}

\usepackage{graphicx}
\usepackage{dcolumn}
\usepackage{bm}
\usepackage{mathrsfs}
\usepackage{subfigure}
\usepackage{natbib}
\usepackage{amsmath}
\usepackage{amssymb}
\usepackage{Jonasmacros}
\usepackage{txfonts}
\usepackage{ifsym}

\usepackage[colorlinks,citecolor=red,linkcolor=Fuchsia]{hyperref}
\usepackage[usenames,dvipsnames,svgnames]{xcolor}

\newcommand{\tobedeleted}[1]{\textcolor{green}{#1}}
\renewcommand{\tobedeleted}[1]{\relax}

\renewcommand{\section}[1]{{\par\it #1.---}\ignorespaces}

\begin{document}
\date{\today} 
\title{Electronically Mediated Magnetic Anisotropy in Vibrating Magnetic Molecules}

\author{J. D. Vasquez Jaramillo}
\author{H. Hammar}
\author{J. Fransson}
\email{Jonas.Fransson@physics.uu.se}
\affiliation{Department of Physics and Astronomy, Box 516, 75120, Uppsala University, Uppsala, Sweden}


\begin{abstract}
We address the electronically induced anisotropy field acting on a spin moment comprised in a vibrating magnetic molecule located in the junction between ferromagnetic metals. Under weak coupling between the electrons and molecular vibrations, the nature of the anisotropy can be changed from favoring a high spin (easy axis) magnetic moment to a low spin (easy plane) by applying a temperature difference or a voltage bias across the junction. For unequal spin-polarizations in the ferromagnetic metals it is shown that the character of the anisotropy is essentially determined by the properties of the weaker ferromagnet. By increasing the temperature in this metal, or introducing a voltage bias, its influence can be suppressed such that the dominant contribution to the anisotropy is interchanged to the stronger ferromagnet. With increasing coupling strength between the molecular vibrations and the electrons, the nature of the anisotropy is locked into favoring easy plane magnetism.
\end{abstract}
\maketitle

Magnetic interactions and anisotropies are fundamentally important for magnetic ordering and magnetic properties of both materials \cite{Jungwirth2016,Batignani2015} and molecular complexes \cite{Science.335.196,Ganzhorn2016,Khajetoorians2010,Wu2017,Grindell2016}.  Recent demonstrations of stabilized ordered magnetic configurations in clusters of magnetic adatoms deposited on metallic surface \cite{Science.335.196}, low frequency switching between degenerate ground states \cite{Science.339.55}, and magnetic remanence of single atoms \cite{Science.352.318}, open for atomic and molecular scale magnetism. Reports of electrical control of the amplitude of the magnetic anisotropy \cite{NanoLett.10.3307,NanoLett.15.4024} show versatile functionality. Moreover, the sign of the anisotropy is fundamentally important for whether the spin assumes a high (easy axis) or a low (easy plane) spin ground state. The general mechanisms that control the sign of the anisotropy remains, nonetheless, an open question \cite{ChemRev.113.5110,ChemComm.50.1648,ChemSci.2.2078,ChemSocRev.40.3092,InorgChem.48.3467,Baumann2015,Srtio2007,Osorio2010}.

Crucial for the demonstrations in, e.g., Refs. \cite{Science.335.196,Science.339.55,Science.352.318,NanoLett.10.3307,NanoLett.15.4024}, is that the effective anisotropy field acting on the local magnetic moments is sufficiently strong to suppress rapid fluctuations between degenerate and nearly degenerate magnetic configurations. However, experiments of this kind is often performed at low temperatures since the energy scale of the pertinent magnetic anisotropies is of the order of sub-meV and, thus, are not effective at high temperatures \cite{Sharples2014,NatNano.8.165,Ganzhorn2016}.

Another aspect of low temperature measurements is that ionic displacements and vibrations can be subdued, thus, quenching undesired configurational fluctuations which may transfer angular momentum into the spin degrees of freedom. It is well-known, however, that molecular vibrations may have a severe influence on the electronic and magnetic properties of molecular structures \cite{Palyi2012,NatNano.8.165,Krainov2017,Kenawy2017,PhysRevLett.116.185501}. The resulting modifications of the internal properties are then conveyed over to, e.g., the transport properties, which has been shown both experimentally \cite{Sohn2017,NatNano.8.165,Ganzhorn2016,Pradip2016} and theoretically \cite{Lassagne2011,PhysRevB.73.045314,JPCM.19.103210,Misiorny2013,Bessis2016,Lunghi2017b,Lunghi2017}.

\begin{figure}[t]
\begin{center}
\includegraphics[width=\columnwidth]{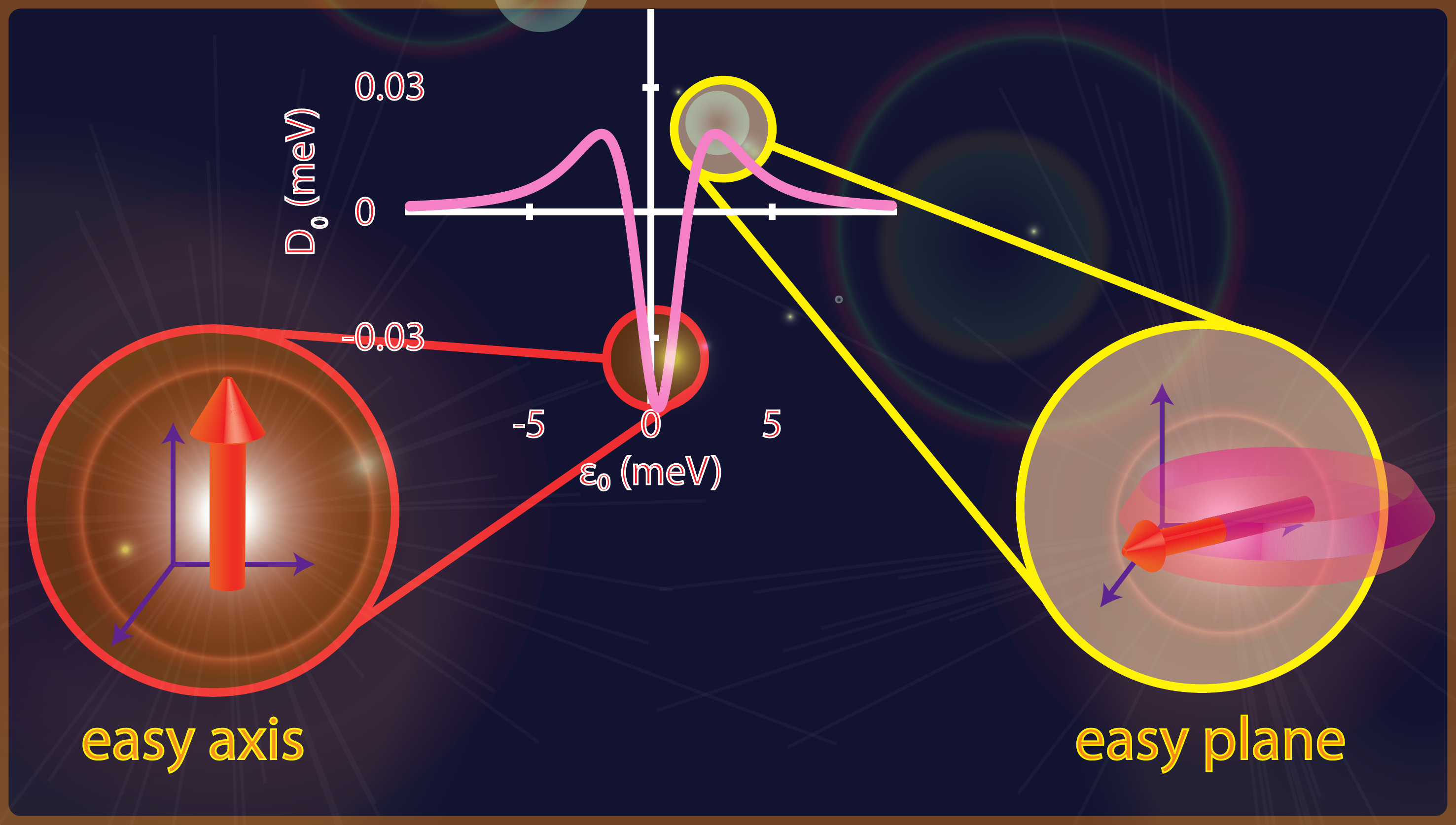}
\end{center}
\end{figure}

Thus far, issues related to ionic displacements and vibrations have been studied separately from questions concerning magnetically relevant quantities in molecular structures \cite{JPCM.19.103210,Jafri2013}. At a phenomenological level, however, there have been several studies in which effective coupling rates between vibrational and spin degrees of freedom have been employed to account for thermal fluctuations through molecular vibrations, see, e.g., Refs. \cite{ChemSci.4.139,NatNano.8.165,NatPhys.11.820,Lunghi2017b}. On the other hand, the coupling rates have been inserted by hand in an \emph{ad hoc} fashion and does not relate to the real microscopic nature of the system. In this Letter, we consider the magnetic anisotropy acting on a local magnetic moment, embedded in a molecular environment, which is induced by the molecular electronic structure. In order to go beyond the simple electronic picture of the anisotropy, we include a weak coupling between the electrons and molecular vibrations. We make contact with a possible experimental set-up by considering the molecular compound to be inserted in the junction between ferromagnetic metallic electrodes, as studied, e.g., in \cite{NatNano.8.165}. It is important to notice that the anisotropy crucially depends on the molecular electronic and magnetic structure near the chemical potential in the set-up but also on the occupied electronic density. The interplay between these components and of the spin-polarizations in the ferromagnets, determine the sign of the anisotropy.
We, furthermore, find that the electronically induced magnetic anisotropy strongly, in addition to the the spin-polarizations of the leads, depends on the temperature difference and voltage bias across the junction. In particular, the sign of the anisotropy field can be deliberately switched, in junctions with different spin-polarizations in the two ferromagnets, by increasing the temperature in the ferromagnet with stronger spin-polarization. Furthermore, when inferring non-equilibrium conditions by applying a voltage bias across the junction, the polarity of the voltage determines the sign of the anisotropy. Of major importance is that our results remain robust under the presence of molecular vibrations weakly coupled to the electrons. This may not be expected since the molecular vibrations tend to broaden the distribution of the electronic density. However, since the contribution associated with the electronic density at the chemical potential can be tuned by external gating, it is feasible to vary both the sign and amplitude by external forces. For sufficiently strong electron-vibron coupling the sign of the anisotropy is locked into favoring easy plane ground state only.

\section{Model}
We consider the magnetic anisotropy in a single magnetic molecule embedded in the junction between ferromagnetic or normal metallic leads. The molecule comprises a ligand structure, which provides the highest occupied molecular orbital (HOMO) or lowest unoccupied molecular orbital (LUMO) close to the Fermi level of the system as a whole, which is connected to the leads via tunneling. Henceforth we shall refer to the HOMO/LUMO level as the localized level. Located in the molecular structure is also a spin moment $\bfS$ which interacts via local exchange (Kondo coupling) with the localized level. Molecular vibrations are, moreover, coupled to the localized level. All in all, the structure is modelled using the Hamiltonian
\begin{align}
\Hamil=&
	\Hamil_L+\Hamil_R+\Hamil_T+\Hamil_\text{mol}
	.
\end{align}
Here, the energies of the metallic leads are given by $\Hamil_\chi=\sum_{\bfk\sigma\in\chi}\leade{\bfk}\cdagger{\bfk}\cc{\bfk}$, where $\chi$ denotes the left ($L$) and right ($R$) leads, in which electrons are created (annihilated) by the operator $\cdagger{\bfk}$ ($\cc{\bfk}$) at the energy $\leade{\bfk}$ with momentum $\bfk$ and spin $\sigma=\up,\down$. For convenience, we denote electrons in the left (right) lead with the momentum $\bfk=\bfp$ ($\bfk=\bfq$).
Tunneling between the leads and the localized level is described in terms of $\Hamil_T=\sum_{\bfk\sigma}t_{\bfk\sigma}\cdagger{\bfk}\dc{\sigma}+H.c.$, where $t_{\bfk\sigma}$ defines the spin-dependent tunneling rate, whereas $\dc{\sigma}$ denotes the electron annihilation in the localized level.
The molecular complex is, finally, described through the Hamiltonian
\begin{align}
\Hamil_\text{mol}=&
	\sum_\sigma
	\Bigl(
		\dote{0}
		+
		Un_{\bar\sigma}/2
		+
		\lambda(b+b^\dagger)
	\Bigr)
	n_\sigma
	-
	v\bfs\cdot\bfS
	+
	\omega_0b^\dagger b
	,
\end{align}
where $\dote{0}$ denotes the spin-independent energy of the localized level, $U$ is the local Coulomb repulsion, and $n_\sigma=\ddagger{\sigma}\dc{\sigma}$. The electron spin $\bfs=\sum_{\sigma\sigma'}\ddagger{\sigma}\bfsigma_{\sigma\sigma'}\dc{\sigma'}/2$ in the localized level interact with the local spin $\bfS$ with rate $v$, where $\bfsigma$ is the vector of Pauli matrices. The local vibrational mode is described by the Bosonic operators $b^\dagger$ and $b$ and energy $\omega_0$, and is coupled to electrons in the localized level with rate $\lambda$.

The coupling between the electronic and vibrational degrees of freedom is considered weak, such that the local electronic structure is only weakly modified by the vibrations. Although this assumption limits the applicability of our conclusions, it nevertheless provides qualitative information about some expected influences from the vibrations on the local magnetic moment. For stronger couplings and more general conclusions one would have to employ self-consistent methods \cite{PhysRevB.73.045314,PhysRevB.78.125320} or coupled spin and lattice dynamics\cite{PhysRevMaterials.1.074404}. However, under the assumption of weak coupling we can employ the Lang-Firsov transformation \cite{JETP.16.1301} which leads to a diagonal formulation of $\Hamil_\text{mol}$ with respect to the effective Fermionic and Bosonic degrees of freedom.

The Lang-Firsov transformation is performed through the canonical transformation $\tilde{\Hamil}=e^\calS\Hamil e^{-\calS}$, with the generating functional $\calS=-(\lambda/\omega_0)(b-b^\dagger)\sum_\sigma n_\sigma$. The transformation leads to that the localized energy levels are shifted into $\tilde{\dote{}}_\sigma-\lambda^2/\omega_0$, while the renormalized charging energy $\tilde{U}=U-2\lambda^2/\omega_0$. The Hamiltonian for the molecular structure is, accordingly, reduced to
\begin{align}
\tilde{\Hamil}_\text{mol}=&
	\sum_\sigma
	\Bigl(
		\tilde{\dote{}}_\sigma
		+
		\tilde{U}n_{\bar\sigma}/2
	\Bigr)
	n_\sigma
	-
	v\bfs\cdot\bfS
	+
	\omega_0b^\dagger b
	.
\end{align}
The tunneling rates in $\Hamil_T$ are changed into $t_{\bfk\sigma}X$, where $X=\exp\{(\lambda/\omega_0)(b-b^\dagger)\}$. However, considering only weakly coupled electrons and vibrations \cite{JPC.13.4477,PhysRevB.73.045314,JPCM.19.103210}, the factor $X\approx1$ such that we can replace the tunneling rates by $t_{\bfk\sigma}$. Hence, the remaining three contributions in the Hamiltonian are considered unaffected by the transformation.

\section{Green function approach}
We calculate physical quantities, such as, the electronically induced magnetic anisotropy acting on the local spin moment and the density of electron states of the localized level in terms of the single electron Green function $\bfG(t,t')=\{\eqgr{\dc{\sigma}(t)}{\ddagger{\sigma'}(t')}\}_{\sigma\sigma'}$, which is a $2\times2$-matrix in spin 1/2 space. For later purposes we note that the Green function can be decomposed into its charge and magnetic components $G_0$ and $\bfG_1$, respectively, according to $\bfG=G_0\sigma^0+\bfG_1\cdot\bfsigma$, where $\sigma^0$ is the identity matrix. Hence, the non-equilibrium density of electron states and the corresponding spin-polarization are obtained from
\begin{subequations}
\begin{align}
\rho(\omega)=&
	\frac{i}{2\pi}
		\tr\Bigl(\bfG^>(\omega)-\bfG^<(\omega)\Bigr)
	=
	\frac{i}{\pi}
		\Bigl(G_0^>(\omega)-G_0^<(\omega)\Bigr)
	,
\\
\bfrho_s(\omega)=&
	\frac{i}{2\pi}
		\tr\bfsigma\Bigl(\bfG^>(\omega)-\bfG^<(\omega)\Bigr)
	=
	\frac{i}{\pi}
		\Bigl(\bfG_1^>(\omega)-\bfG_1^<(\omega)\Bigr)
	,
\end{align}
\end{subequations}
respectively.

Given that the leads are aligned either in parallel or anti-parallel to one another, we can introduce a global spin-quantization axis, say, $\hat{\bf z}$, along which all magnetic properties are given. It is, thus, justified to assume that $\bfG_1=G_z\hat{\bf z}$, such that we can define $G_{0(z)}=\sum\sigma^{0(z)}_{\sigma\sigma}G_\sigma/2$, where $G_\sigma$ is the spin projected Green function. Finally we notice that since we are interested in non-equilibrium properties, we calculate all quantities in terms of the lesser and greater propagators $\bfG^{</>}$. For simplicity, however, and since we view the leads as source and drain reservoirs for electrons, the lesser/greater Green functions will be calculated using the relation $\bfG^{</>}=\bfG^r\bfSigma^{</>}\bfG^a$, where $\bfG^{r/a}$ is the retarded/advanced forms of the Green function, whereas $\bfSigma^{</>}$ is the lesser/greater self-energy. Here, we shall merely include the couplings $\bfGamma^\chi$ to the reservoirs in this self-energy, such that we can write $\bfSigma^{</>}(\omega)=(\pm i)\sum_\chi f_\chi(\pm\omega)\bfGamma^\chi$, where $f_\chi(\omega)=f(\omega-\mu_\chi)$ is the Fermi distribution function for the electrons in the lead $\chi$ with (electro-) chemical potential $\mu_\chi$. The spin-resolved coupling parameters $\Gamma_\sigma^\chi=2\pi\sum_{\bfk\in\chi}|t_{\bfk\sigma}|^2\delta(\omega-\leade{\bfk})$ are parametrized such that $\Gamma^\chi_\sigma=\Gamma(1+\sigma^z_{\sigma\sigma}p_\chi)/4$, $-1\leq p_\chi\leq1$, where $\Gamma=\sum_{\chi\sigma}\Gamma^\chi_\sigma$, $\Gamma^\chi=\sum_\sigma\Gamma^\chi_\sigma$, and $\Gamma_\sigma=\sum_\chi\Gamma^\chi_\sigma$.

From the Lang-Firsov transformation it also follows that the single electron Green function
\begin{align}
G_{\sigma\sigma'}(t,t')=&
	\eqgr{(\dc{\sigma}X)(t)}{(\ddagger{\sigma'}X^\dagger)(t')}
\nonumber\\=&
	\tilde{G}_{\sigma\sigma'}(t,t')\av{X(t)X^\dagger(t')}
	,
\end{align}
where the propagators $\tilde{G}_{\sigma\sigma'}(t,t')=\eqgr{\tilde{\dc{}}_\sigma(t)}{\tilde{\ddagger{}}_{\sigma'}(t')}$ and $\av{X(t)X^\dagger(t')}$ are defined in the model $\tilde{\Hamil}$ such that $\tilde{d}_\sigma(t)=\exp\{i\tilde{\Hamil}_\text{electron}t\}\tilde{d}_\sigma\exp\{-i\tilde{\Hamil}_\text{electron}t\}$ and $X(t)=\exp\{i\tilde{\Hamil}_\text{vibration}t\}X\exp\{-i\tilde{\Hamil}_\text{vibration}t\}$. The renormalization factor caused by the coupling between the electrons and vibrations is calculated through $\av{X(t)X^\dagger(t')}=\exp\{-\Phi(\tau)\}$, where \cite{ManyParticlePhysics} ($\tau=t-t'$)
\begin{align}
\Phi(\tau)=&
	\biggl(\frac{\lambda}{\omega_0}\biggr)^2
	\Bigl(
		n_B
		\Bigl[
			1-e^{i\omega_0\tau}
		\Bigr]
		+
		(1+n_B)
		\Bigl[
			1-e^{-i\omega_0\tau}
		\Bigr]
	\Bigr)
	.
\end{align}
Here, $n_B=(\exp\{\beta\omega_0\}-1)^{-1}$ is the Bose-Einstein distribution with the inverse temperature $\beta=1/k_BT$ in terms of the Boltzmann constant $k_B$. Furthermore, we calculate the electronic propagator $\tilde\bfG$ in the mean-field approximation, where the retarded components are given by $\tilde{G}^r_{\sigma\sigma'}=\delta_{\sigma\sigma'}\tilde{G}^r_\sigma$, where
\begin{align}
\tilde{G}^r_\sigma(\omega)=&
	\frac{\omega-\tilde{\dote{}}_\sigma-(1-\av{n_{\bar\sigma}})\tilde{U}}
	{
	(\omega-\tilde{\dote{}}_\sigma+i\Gamma_\sigma/2)(\omega-\tilde{\dote{}}_\sigma-\tilde{U})
	+
	i\av{n_{\bar\sigma}}\tilde{U}\Gamma_\sigma/2}
	,
\end{align}
and $\av{n_\sigma}=(-i)\int\tilde{G}^<_\sigma(\omega)d\omega/2\pi$. It is, then, a straight forward calculation to obtain
\begin{align}
G^r_\sigma(\omega)=&
	e^{-(\lambda/\omega_0)^2(2n_B+1)}
	\sum_n
		I_n(z)e^{n\beta\omega_0/2}
		\tilde{G}^r_\sigma(\omega-n\omega_0)
	,
\end{align}
where $I_n(z)$ is the $n$th modified Bessel function and $z=2(\lambda/\omega_0)^2\sqrt{n_B(n_B+1)}$.

We infer the limitations of the present approximations \cite{JPC.13.4477} on $\lambda$, $\omega_{0}$, and $\Gamma$, given by $\lambda/\omega_0<1<(\Gamma/2\omega_0)\exp\{-(\lambda/\omega_{0})^2\}$, $\omega_0<|\epsilon_\sigma-\lambda^2/\omega_0|$, and $2\lambda^2/\omega_0<\Gamma$. In order to fulfill these restrictions we require that $\lambda/\Gamma\leq1/6$ and $1/5<\omega_0/\Gamma<6/5$.

\section{Anisotropy parameter}
The influence of the electronic motion on the localized spin moment $\bfS$ have been considered previously in studies of, e.g., the spin and transport dynamics in single magnetic molecules \cite{PhysRevB.94.054311,PhysRevB.96.214401}, and the interplay between magnetic interactions, configurations, and transport in dimers of magnetic molecules \cite{PhysRevLett.113.257201,NanoLett.16.2824,JPhysChemC.121.27357}, as well as in general materials structures \cite{PhysRevMaterials.1.074404}. Following the prescription formulated in these publications, we here write the electronically induced magnetic anisotropy acting on the localized spin moment according to \cite{PhysRevLett.113.257201}
\begin{align}
I_{zz}=&
	v^2
	\int
		\frac{G_z^>(\omega)G_z^<(\omega')-G_z^<(\omega)G_z^>(\omega')}{\omega-\omega'}
	\frac{d\omega}{2\pi}
	\frac{d\omega'}{2\pi}
	.
\end{align}
While the anisotropy is generally formulated as a second rank tensor, it is here reduced to the $zz$-component only, since all magnetic properties are aligned with the global $\hat{\bf z}$-direction. In the framework of an effective model for the localized spin, where the surrounding environment is parametrized into local interactions and anisotropies \cite{PhysRevMaterials.1.074404}, we write the spin Hamiltonian as
\begin{align}
\Hamil_S=&
	DS_z^2
	+
	E(S_x^2-S_y^2)
	,
\end{align}
where $D$ and $E$ are the uniaxial and transverse anisotropy parameters, respectively. In the present set-up, where the spin direction is given along $\hat{\bf z}$, the transverse anisotropy $E=0$ whereas the uniaxial anisotropy $D\equiv I_{zz}$.

Within the scheme presented above, we can derive a simplified expression for the anisotropy parameter $D$ in terms of the retarded Green function for the localized level and the coupling parameters to the leads. After some straightforward algebra, we obtain
\begin{align}
D=&
	-\frac{v^2}{4}
	\sum_\chi
	\sum_{\sigma\sigma'}
		\sigma^z_{\sigma\sigma}\sigma^z_{\sigma'\sigma'}
		\Gamma_\sigma^\chi
		\Gamma_{\sigma'}
\nonumber\\&\times
		\int
			f_\chi(\omega)
			|G^r_\sigma(\omega)|^2
			\frac{|G^r_{\sigma'}(\omega')|^2}{\omega-\omega'}
		\frac{d\omega}{2\pi}
		\frac{d\omega'}{2\pi}
	.
\end{align}
Here, in the absence of both the coupling between the electrons and vibrations ($\lambda=0$), as well as, the Coulomb interaction ($U=0$), the current approximations allow to make the identification $\Gamma_\sigma|G_\sigma^r(\omega)|^2/2=-\im\, G^r_\sigma(\omega)$. Thence, using the Kramers-Kr\"onig relations, it is straight forward to see that the \emph{bare} anisotropy parameter $D_0$ can be written according to
\begin{align}
D_0=&
	-\frac{v^2}{2}
	\re
	\sum_\chi
	\sum_{\sigma\sigma'}
		\sigma^z_{\sigma\sigma}\sigma^z_{\sigma'\sigma'}
		\Gamma^\chi_\sigma
		\int
			f_\chi(\omega)
			|G^r_\sigma(\omega)|^2
			G^r_{\sigma'}(\omega)
		\frac{d\omega}{2\pi}
\nonumber\\=&
	D_0^\sim
	+
	D_0^\top
	,
\end{align}
where we have introduced the components
\begin{subequations}
\label{eq-D0contributions}
\begin{align}
D_0^\sim=&
	\frac{v^2}{2}
	\im
	\sum_\chi
		\int
			\frac{\partial f_\chi(\omega)}{\partial\omega}
			\Biggl(
				\frac{\Gamma^\chi_\up}{\Gamma_\up}
				G^r_\up(\omega)
				-
				\frac{\Gamma^\chi_\down}{\Gamma_\down}
				G^r_\down(\omega)
			\Biggr)
\nonumber\\&\hspace{2cm}\times
			\ln\Biggl|\frac{G^r_\up(\omega)}{G^r_\down(\omega)}\Biggr|
		\frac{d\omega}{2\pi}
	,
\label{eq-D0surface}
\\
D_0^\top=&
	-\frac{v^2}{2}
	\im
	\sum_\chi
		\int
			f_\chi(\omega)
			\Biggl(
				\frac{\Gamma^\chi_\up}{\Gamma_\up}
				\Bigl(G^r_\up(\omega)\Bigr)^2
				-
				\frac{\Gamma^\chi_\down}{\Gamma_\down}
				\Bigl(G^r_\down(\omega)\Bigr)^2
			\Biggr)
\nonumber\\&\hspace{2cm}\times
			\ln\Biggl|\frac{G^r_\up(\omega)}{G^r_\down(\omega)}\Biggr|
		\frac{d\omega}{2\pi}
	.
\label{eq-D0sea}
\end{align}
\end{subequations}
From this expression it can be understood that the anisotropy crucially depends on the spin-imbalance ($\ln|G_\up^r|-\ln|G^r_\down|$) in the molecular electronic structure. The specific details of the electronic structure near the chemical potentials $\mu_\chi$ of the leads is provided through the contribution $D_0^\sim$ since it involves the derivative $-\partial f_\chi(\omega)/\partial \omega$. The strength as well as the sign of the anisotropy energy is determined by the interplay between the contributions $D_0^\sim$ and $D_0^\top$, manifest through their opposite signs. The former component is large whenever there is a simultaneous large local electron density and large local spin-imbalance at the chemical potentials $\mu_\chi$. This follows from the presence of the derivatives of the Fermi function, which pick out the value of the spin-resolved density of electron states, $-\im G^r_\sigma$, at these energies. The amplitude of the latter component ($D_0^\top$) depends, in addition to the magnitude of the spin-imbalance, on the electronic occupation. Trivially, this component vanishes identically when the electronic density is completely empty, that is, in the limit $\mu_\chi\rightarrow-\infty$, for all $\chi$. In opposite limit, however, there may be a finite value caused by, e.g., distinct occupation numbers of the molecular spin states. The amplitude of the contribution $D_0^\top$ is, nevertheless, maximal when the localized level is resonant with the chemical potentials.
Henceforth, we shall refer to $D_0^\sim$ as Fermi surface effect and $D_0^\top$ as the volume, or, Fermi sea effect -- in correspondence to a terminology often employed in condensed matter physics. We thereby allude to their intrinsic dependences on the electronic structure at the chemical potentials and on the occupied electron density, respectively.

\section{Zero voltage bias}
\begin{figure}[t]
\begin{center}
\includegraphics[width=0.99\columnwidth]{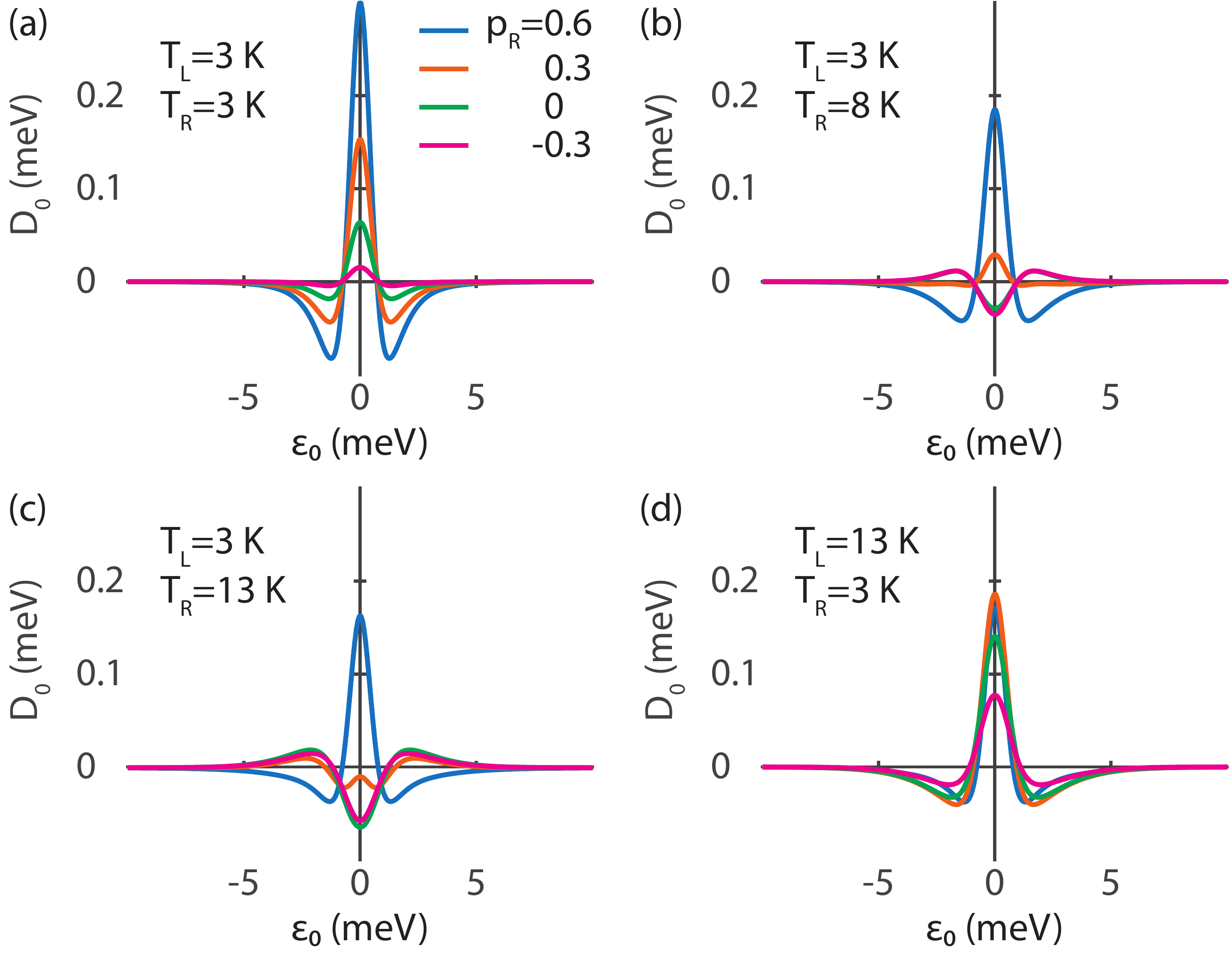}
\end{center}
\caption{Anisotropy as function of the localized energy level $\dote{0}$ for different spin-polarizations $p_R=0.6,\ 0.3,\ 0,\ -0.3$ at constant $p_L=0.6$, and different temperatures in the right lead (a) $T_R=3$ K, (b) $T_R=8$ K, and (c) $T_R=13$ K, at constant temperature $T_L=3$ K in the left, and (d) $T_L=13$ K and $T_R=3$ K. Other parameters are $\Gamma=4$ meV, $\lambda=0$, $U=0$, and $\mu_\chi=0$.}
\label{fig-D0}
\end{figure}

The anisotropy has a non-trivial dependence on the spin-polarization both in the leads and locally in the molecular structure, as well as on the external conditions which may, or may not, plunge the system into non-equilibrium. However, in a simple set-up with a spin-degenerate localized level, where the only spin-dependence is provided from the spin-dependent coupling parameters $\Gamma^\chi_\sigma$, the anisotropy is expected to be positive and large near resonance ($\dote{0}-\mu_\chi\approx0$) while it is negative and small away from resonance. This behavior is verified in Fig. \ref{fig-D0} (a), where we plot the $D_0$ as function of the localized level $\dote{0}$ with respect to the equilibrium chemical potential $\mu=0$ ($\mu_\chi=\mu$). In the simulations we assumed $p_L=0.6$ and $p_R\in\{0.6,\ 0.3,\ 0,\ -0.3\}$, $T_{L/R}=3$ K. The behavior of the anisotropy, varying between negative and positive, opens for the possibility to switch the localized spin moment between high and low spin configurations, where a positive (negative) anisotropy lead to a low (high) spin ground state.

Upon increasing the temperature in the right lead, the influence of the thermal broadening becomes asymmetric, something that has a dramatic influence on the resulting anisotropy. In Fig. \ref{fig-D0} (b), (c), we plot the corresponding $D_0$ for two finite temperature differences (5 K and 10 K, respectively) at different spin-polarizations $p_R$. The most notable feature is that the sign of $D_0$ changes in all set-ups with asymmetric spin-polarzations ($p_L\neq p_R$) when the temperature difference increases.

\begin{figure}[t]
\begin{center}
\includegraphics[width=0.99\columnwidth]{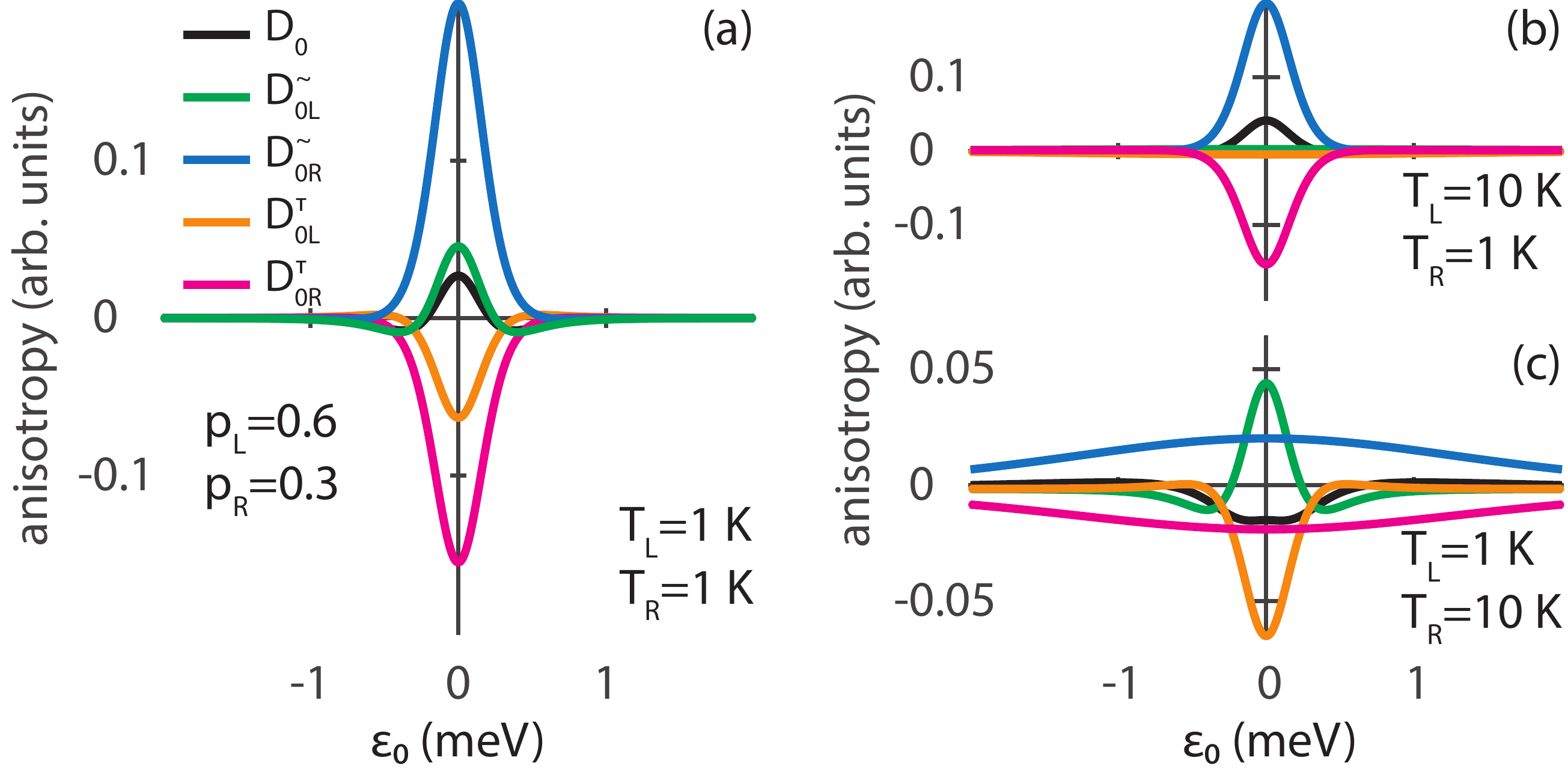}
\end{center}
\caption{Anisotropy (black) along with its left and right Fermi surface ($D_{0L}^\sim$ -- cyan, $D_{0R}^\sim$ -- blue) and Fermi sea ($D_{0L}^\top$ -- orange, $D_{0R}^\top$ -- magenta) components as function of the localized energy level $\dote{0}$ for the spin-polarizations $p_L=2p_R=0.6$ at the temperatures (a) $T_L=T_R=1$ K, (b) $T_L=10T_R=10$ K, and (c) $T_L=T_R/10=1$ K. Other parameters are as in Fig. \ref{fig-D0}.}
\label{fig-D0contributions}
\end{figure}

We analyze this property by resolving the Fermi surface and Fermi sea effects into left and right components according to $D_0^{\sim/\top}=D_{0L}^{\sim/\top}+D_{0R}^{\sim/\top}$, which is a natural partitioning given the summation over the left and right electrodes in Eq. (\ref{eq-D0contributions}). In Fig. \ref{fig-D0contributions} (a) the total anisotropy $D_0$ (black) is plotted along with its components (see figure for details), for spin-polarizations $p_L=2p_R=0.6$ and vanishing temperature difference between the leads. It is clear from the plots that the left and right contributions are strongly asymmetric and that the right components dominate the total anisotropy, in this case. Here, $D_{0R}^\sim>0$ and $D_{0R}^\top<0$ near resonance and since the amplitude of the former is the larger one, the result is $D_0>0$. This condition remains true under an increase of the temperature $T_L$, see Fig. \ref{fig-D0contributions} (b), since the increased thermal broadening of the components $D_{0L}^\sim>0$ and $D_{0L}^\top<0$ induced from the left lead tends to equalize the sizes of their amplitudes such that $D_{0L}^\sim+D_{0L}^\top\approx0$. However, by increasing the temperature $T_R$, on the other hand, likewise the increased thermal broadening tends to equalize the amplitudes of $D_{0R}^\sim>0$ and $D_{0R}^\top<0$ instead, such that the overall effect from the right lead vanishes, see Fig. \ref{fig-D0contributions} (c). Hence, the properties induced from the left lead begins to dominate and in the set-up depicted in Fig. \ref{fig-D0contributions}, the negative amplitude $D_{0L}^\sim$ is larger than the positive amplitude of $D_{0L}^\top$, which leads to an overall negative anisotropy.

Furthermore, consider the simplified expression of the surface component to the anisotropy in Eq. (\ref{eq-D0surface}), at $\lambda=0$, $U=0$, and let the temperature $T_\chi\rightarrow0$, such that energy derivative of the Fermi function can be replaced by the Dirac delta function $\delta(\omega-\mu_\chi)$. Then, the value of the integrand is picked out at the chemical potential $\mu_\chi$, which enables an analysis of the voltage dependent sign of the anisotropy at $\mu_\chi$. In fact, it is straightforward to show that this contribution is positive whenever the spin-polarization $p_\chi$ satisfies the condition
\begin{align}
|p_\chi|>&
	4\frac{|p_L+p_R|}{4+(p_L+p_R)^2}
	.
\end{align}
Using this relation gives a tool to design and engineer the structure in terms of the ferromagnetic leads such that the nature, or, sign of the anisotropy can be predicted.

The conclusion regarding the anisotropy at zero voltage bias is, hence, that its sign can be switched under the application of a thermal difference across the molecular junction, provided that the spin-polarizations in the leads are unequal. This feature is strongly enhanced the stronger the asymmetry of the spin-polarizations of the leads is along with a large temperature difference between the leads. The properties of the anisotropy is dominated by the properties of the lead with the lowest temperature. If both temperatures are equal or nearly equal, the lead with the weaker spin-polarization has the strongest influence. In the asymmetric set-up, the stronger spin-polarization tends to generate a negative anisotropy near resonance, while the weaker generates a positive. Since the latter typically has the larger amplitude, the overall anisotropy becomes positive. By increasing the temperature in the lead with weaker (stronger) spin-polarization, the effect of the lead with the stronger (weaker) becomes increasingly important.

\begin{figure}[t]
\begin{center}
\includegraphics[width=0.99\columnwidth]{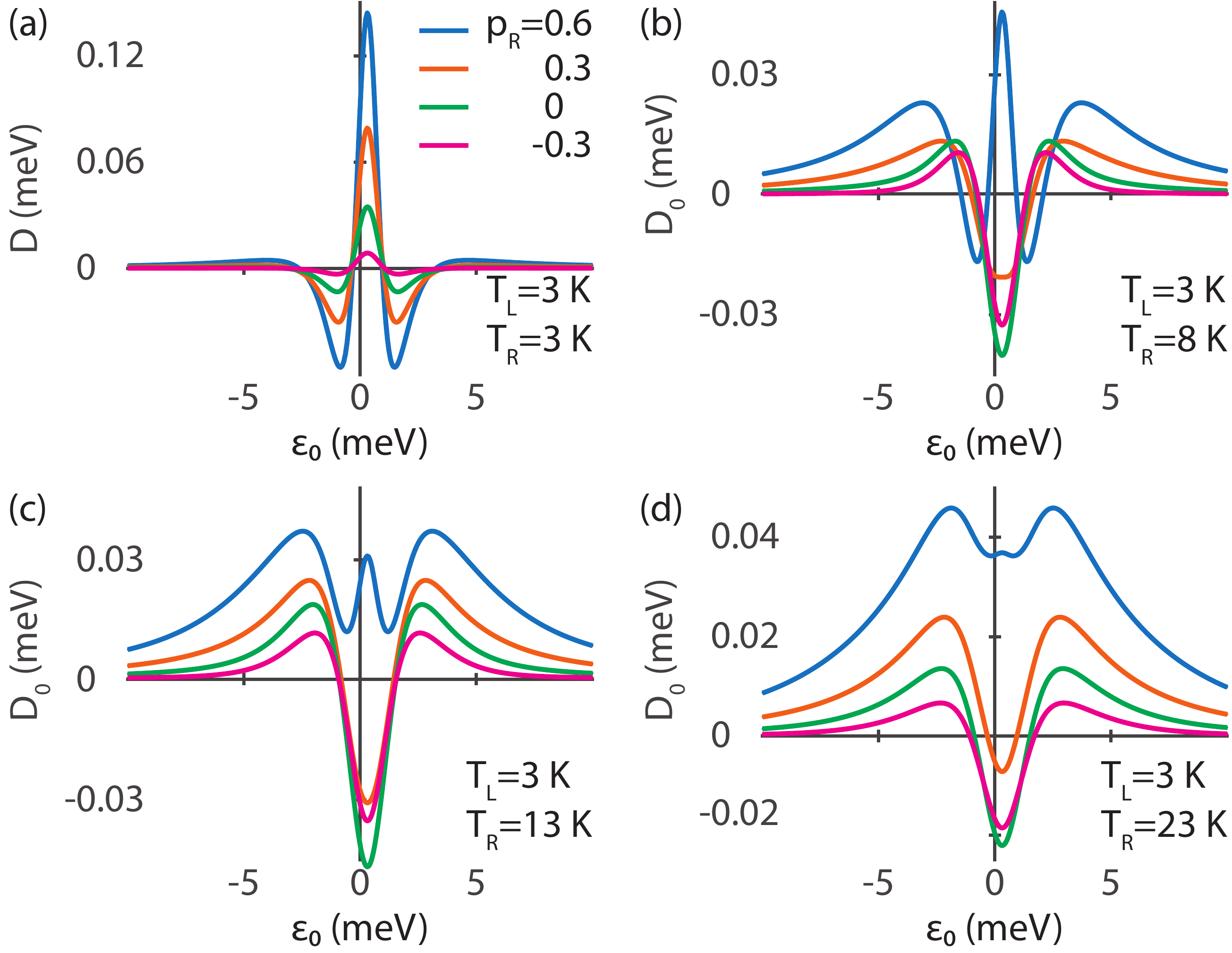}
\end{center}
\caption{Anisotropy as function of the localized energy level $\dote{0}$ for different spin-polarizations $p_R=0.6,\ 0.3,\ 0,\ -0.3$ at constant $p_L=0.6$, and different temperatures in the right lead (a) $T_R=3$ K, (b) $T_R=8$ K, (c) $T_R=13$, and (d) $T_R=23$ K, at constant temperature $T_L=3$ K in the left, $\lambda=\Gamma/8$, and $U=2\lambda^2/\omega_0$. Other parameters are as in Fig. \ref{fig-D0}.}
\label{fig-D}
\end{figure}
\begin{figure}[t]
\begin{center}
\includegraphics[width=0.99\columnwidth]{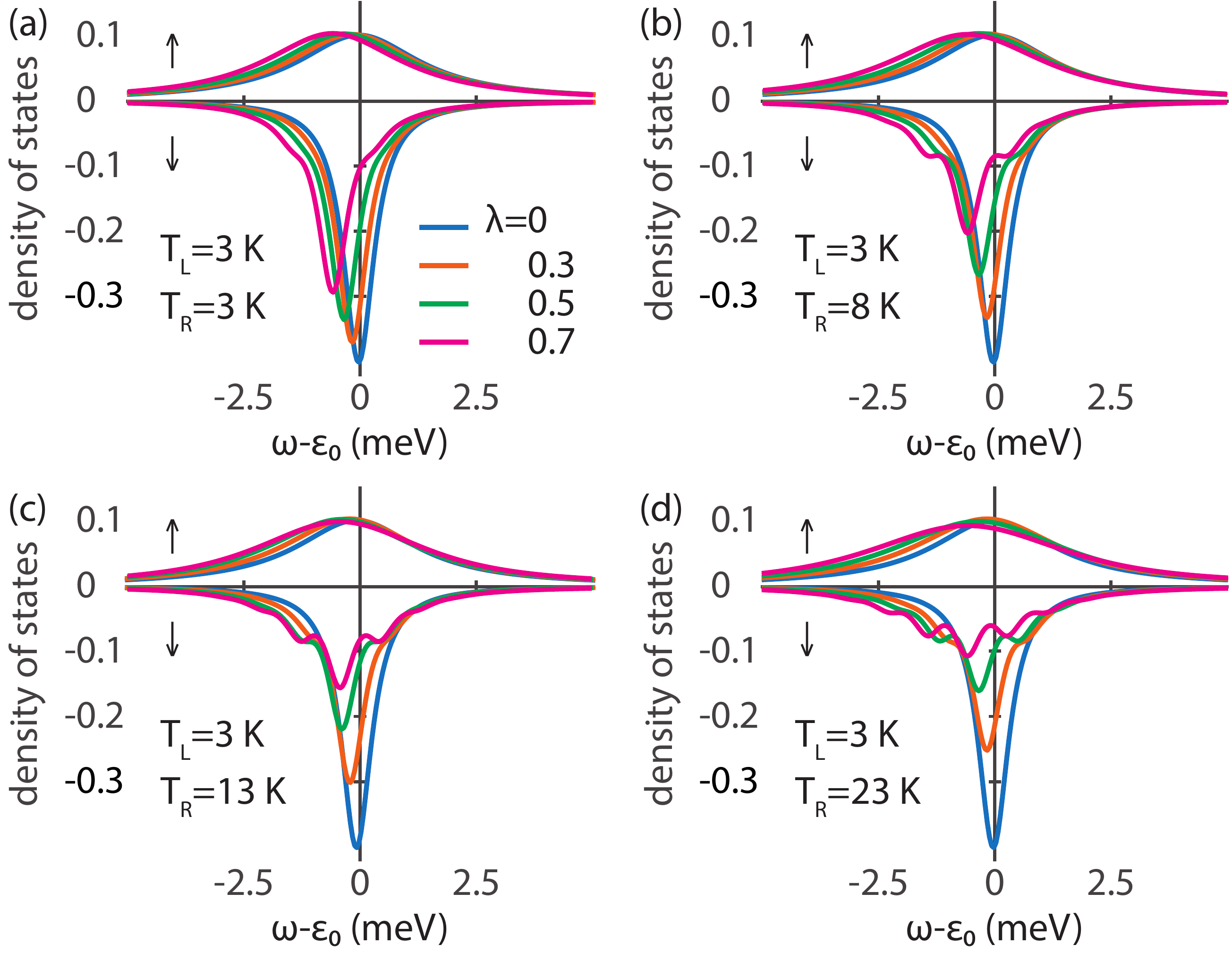}
\end{center}
\caption{Spin-resolved densities of electron states for different couplings $\lambda/\Gamma=0,\ 1/12,\ 1/8,\ 1/6$  at constant $p_L=2p_R=0.6$ and different temperatures in the right lead (a) $T_R=3$ K, (b) $T_R=8$ K, (c) $T_R=13$ K, and (d) $T_R=23$ K, at constant temperature $T_L=3$ K in the left. Other parameters are in Fig. \ref{fig-D}.}
\label{fig-densities}
\end{figure}

A compelling issue, regarding the sign changes of the anisotropy is whether these features remain under the influence of molecular vibrations. In particular, one concern might be that the sign change at resonance is suppressed by the broadening of the electron density caused when vibrational fluctuations modify the electronic structure. However, by investigating the anisotropy in presence of the molecular vibrations ($\lambda>0$), we find that most of the features that were obtained in absence of the vibrations are quite robust in the sense that these are retained also when the coupling to the vibrations is finite. In Fig. \ref{fig-D} we plot the anisotropy for the spin-polarizations and temperature differences used in Fig. \ref{fig-D0}. The most noticeable difference compared to the case without molecular vibrations is the general decrease in amplitude near resonance. Associated with this decrease is also a slight broadening of the anisotropy around resonance. We attribute these features to the emergence of several resonances in the electronic density caused by the coupling between the electrons and vibrations, something which is visualized in Fig. \ref{fig-densities}. In this figure we have plotted the spin-resolved densities of electron states for varying coupling $\lambda$ and temperature difference across the junction. In particular, the emergence of vibrational side resonances with increasing $\lambda$ and temperature difference is clearly seen in the spin channel corresponding to smaller couplings to the leads. The energy shift of the resonant conditions, both in the anisotropy and the densities of electron states, seen for finite $\lambda$, is due to the energy renormalization $\tilde{\dote{}}_\sigma=\dote{\sigma}-\lambda^2/\omega_0$ invoked by the Lang-Firsov transformation.

These observations for the anisotropy at zero voltage bias and finite temperature difference solidify our previous conclusion in that the possibility to switch the anisotropy between negative and positive values near resonance remains essentially unaffected by the presence of molecular vibrations. We notice, nonetheless, that while the lowered amplitude of the anisotropy decreases the effective temperature range in which this anisotropy has a viable effect on the local spin moment, its increased broadening around resonance makes it less susceptible to fluctuations in the molecular environment. It is also important to notice that the coupling to vibrations has a tendency to lock the sign of the anisotropy, which striking in the case of equal spin-polarizations in the leads. We discuss this effect in more detail below.

\section{Finite voltage bias}
\begin{figure}[t]
\begin{center}
\includegraphics[width=0.99\columnwidth]{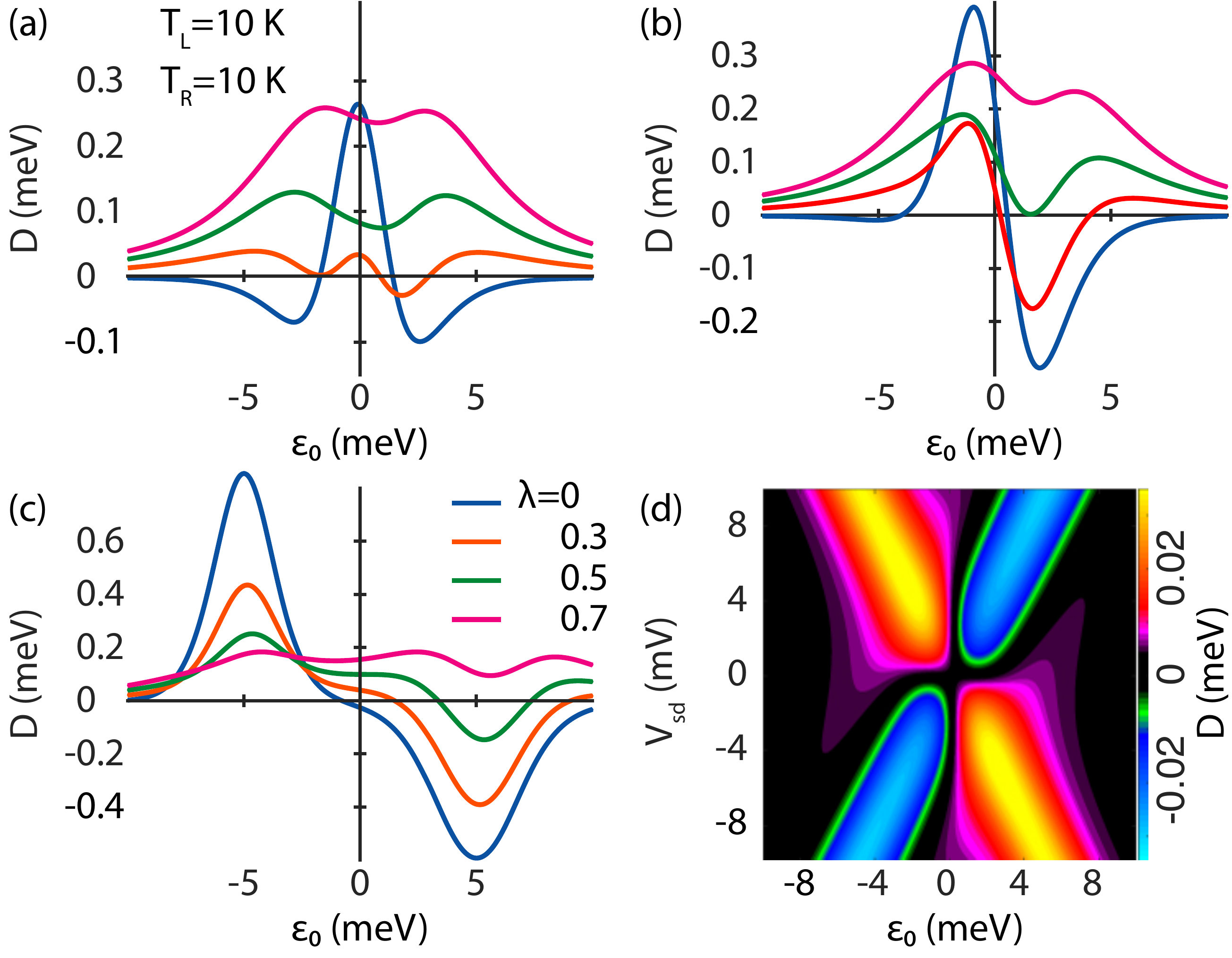}
\end{center}
\caption{Anisotropy as function of the localized energy level $\dote{0}$ for the spin-polarizations $p_L=2p_R=0.6$, temperatures $T_L=T_R=10$ K, and different $\lambda/\Gamma=0,\ 1/12,\ 1/8,\ 1/6$ for voltage biases (a) $V_\text{sd}=0.1$ mV, (b) $V_\text{sd}=1.1$ mV, (c) $V_\text{sd}=10$ mV, and (d) for varying $V_\text{sd}$ and $\lambda=1/10$. Other parameters are as in Fig. \ref{fig-D}.}
\label{fig-Dneq}
\end{figure}

At zero bias, the amplitude and sign of the anisotropy field results from an non-trivial interplay between the Fermi surface and volume effects, on the one hand, and the relation between the spin-polarizations in the two leads, on the other. Under finite temperature difference the latter led to a ceased net contribution to the anisotropy from the lead with the higher temperature. Under finite bias voltages, the situation is somewhat analogous, in the sense that the anisotropy is dominated by the properties, e.g., spin-polarization and temperature, of the lead that is resonant with the molecular level, while the portion emerging from the other lead is negligible. The result of this behavior is that the anisotropy can become positive for one voltage bias polarity and negative for the opposite, something that particularly may occur under asymmetric spin-polarizations in the leads. This prediction is manifest in Fig. \ref{fig-Dneq}, where the anisotropy is shown as function of the molecular level position for different couplings $\lambda$ and voltage biases, see figure caption for details.

Low voltage biases only slightly alter the shape and nature of the anisotropy from its zero voltage bias characteristics, see Fig. \ref{fig-Dneq} (a), which would be expected from a linear response consideration. At higher voltage biases, however, the individual properties of the ferromagnetic leads become legible, see Figs \ref{fig-Dneq} (b), (c), in the signatures of the anisotropy. This behavior underscores that it is the electronic density and spin-polarization, as well as their corresponding occupations, near the chemical potential that determines the nature of the anisotropy. In the situations displayed in Fig. \ref{fig-Dneq}, the left lead has the stronger spin-polarization, $|p_L|>|p_R|$, and following the results from zero voltage bias, one would expect that the influence from this lead is negative and weaker than the one from the right. A positive voltage bias shifts the left (right) chemical potential $\mu_L$ ($\mu_R$) to a higher (lower) energy by $eV/2>0$ $(-eV/2<0$) which leads to different resonant conditions between the molecular level $\dote{0}$ and the left and right chemical potentials. As $\dote{0}$ approaches $\mu_L$ ($\mu_R$), only the spin-polarization in the left (right) lead matters to the resulting anisotropy, which in the present case becomes negative (positive). That the same, but opposite, characteristics is obtained for negative polarity of the voltage bias can be seen in Fig. \ref{fig-Dneq} (d), which shows the anisotropy as function of the molecular level energy and voltage bias. Finally, increasing the strength of the coupling $\lambda$ tends to lock the sign of the anisotropy to favor the low spin ground state, analogous to the situation at zero voltage bias. The vibrationally induced broadening of the density of electron states as well as lowering its amplitude decrease the electron occupation which quenches the volume contribution and opens for the surface contribution to become dominant. Hence, the net anisotropy assumes a positive sign for a wide range of voltage biases.

\section{Conclusions}
We have studied the electronically induced magnetic anisotropy acting on a localized moment embedded in a molecular structure and placed in the junction between ferromagnetic leads. The spin-polarization of the leads allow for a uniaxial anisotropy that can result in either a high (easy axis) or low (easy plane) spin ground state, depending on the sign of the anisotropy. At zero voltage bias and for unequal spin-polarizations in the leads, we notice that the lead with the weaker spin-polarization tends to have the stronger influence on the resulting sign of the anisotropy. By influence of a temperature difference across the junction, the anisotropy can, however, change sign provided that it is the temperature of the lead with lower spin-polarizaton that is increased. Under finite voltage bias and unequal spin-polarizations, the sign of the anisotropy changes upon reversal of the polarity of the voltage bias. Our results are shown to be robust under the influence of molecular vibrations weakly coupled to the electrons. An increasing coupling strength between the electrons and molecular vibrations, tends to lock the sign of the anisotropy to become positive and, hence, favor a low spin (easy plane) ground state configuration.

\section{Acknowledgments}
We thank O. Eriksson, L. Nordstr\"om, P. Oppeneer, and M. Pereiro for fruitful discussions. Financial support from Colciencias (Colombian Administrative Department of Science, Technology and Innovation) and Vetenskapsr\aa det is acknowledged.

\bibliography{SMManisotropy}

\end{document}